\documentclass[twocolumn]{jpsj2}

\usepackage{amsmath, amsfonts, amsthm, amssymb}
\usepackage{graphicx}
\usepackage{bm}
\usepackage{cite}
\usepackage{enumerate}

\def\runauthor{\textsc{H.~Shinaoka} \textit{et al.}}
\makeatletter
\def\@typeset{}
\makeatother
\title{Ill-Contact Effects of d-Orbital Channels in Nanometer-Scale Conductor \\
}
\author{Hiroshi Shinaoka$^{1}$, Takeo Hoshi$^{2,4}$, and Takeo Fujiwara$^{3,4}$}
\inst{$^{1}$ Department of Applied Physics, The University of Tokyo, Tokyo 113-8656, Japan\\
$^{2}$ Department of Applied Mathematics and Physics, Tottori University, Tottori 680-8550, Japan\\
$^{3}$ Center for Research and Development of Higher Education, The University of Tokyo, Tokyo 113-0033, Japan\\
$^{4}$ Core Research for Evolutional Science and Technology  (CREST-JST), Japan Science and Technology Agency, Japan} 

\abst{
Electronic current in a nanometer-size rod is theoretically investigated 
by an  eigen-channel decomposition method 
in nonequilibrium Green's function formalism.
Physical properties, such as the local density of electrons and local current, 
are decomposed into contributions of eigen-channels. 
We observe that the evanescent modes and nonlinear conductance are 
enhanced in d-orbital systems, and 
the structure of the transmission function, local current density, and penetration depth 
are discussed. 
The two effects of the ill-contact at electrodes in d-orbital systems, 
evanescent modes and the nonlinearity of conductance, 
are regarded as originating in the peak structure of the transmission function 
of eigen-channels in the energy region between chemical potentials 
of left and right lead wires.  
}

\kword{metallic nanowire, nonlinear conductance, evanescent wave, ill-contact, nonequilibrium Green's function}
\begin{document}
\maketitle

\section{Introduction}

Nonequilibrium current through atomic-scale contact is very important in 
nanoscience and nanotechnology, 
where we must know in detail the effects of electrodes. 
Electrodes themselves can be scatterers for electron transport 
even if atomic structures are perfectly ordered at electrodes. 
A decay mode is found in electronic current in nanometer-scale conductance, 
when the conductor has freedoms of scattering path at each layer. 
For example, a decay mode is investigated in a monosite chain model 
with multiple orbitals for each site.~\cite{FUKUYAMA} 
The decay mode, evanescent wave, is known also in optics 
as a nonpropagating mode without dissipation 
and is of great interest for a possible foundation of nanophotonics. 
It appears in a nanometer-scale region of material surface, 
when the light is totally reflected at the surface.~\cite{OPTICS}
Realistic nanometer-scale conductors, on the other hand, 
have been studied extensively 
for their unique structural and transport properties.~\cite{agrait03} 
One can find, for example,  quantized conductance in a monatomic wire
through a ballistic channel and resonant conductance 
through a quantized state with strong scattering at electrodes.
In nanometer-scale systems, an atomic structure is crucially important 
for transport properties.~\cite{BETTINI}
The structural investigation is essentially important 
for discussing quantitative properties of realistic conductance of nanoscale rods.  
For example, gold appears as a helical multishell nanowire,~\cite{TAKAYANAGI2000}
whose formation was explained with a two-stage process 
in nanometer-scale phenomena.~\cite{IGUCHI2007}
We like to discuss, in this paper, the characteristic properties 
generic in several nanometer-scale rods. 
Particularly, the possible effect of ill-contact at electrodes 
in nanometer-scale conductors is investigated, 
as a combined effect of structural and transport properties.
The evanescent wave current appears, 
when we carry out a channel-decomposition analysis 
with nonequilibrium Green functions, and 
the nonlinear conductance under a finite bias voltage is observed. 
We will present, in this paper, calculations of 
an electronic structure, for d-orbital systems of rodlike nanowires 
of different lengths and widths,
which gives prototypical cases for real materials. 
The generic ill-contact in the electrodes is enhanced in d-orbital systems 
and becomes an origin of evanescent waves and nonlinear conductance.  
This paper is organized as follows; 
In {\S}~\ref{SEC:THEORY}, we review briefly the formalism of the nonequilibrium 
Green's function and present the eigen-channel decomposition of Green's function, 
local current and other physical quantities. 
We show several calculated results on the decaying properties of 
electron density and the nonlinearity of $I/V$ curves in {\S}~\ref{SEC:TOTAL-DECAY-BEHAVIOR}. 
Section \ref{origin} is devoted to the analysis and discussion of the origin 
of nonlinear conductance and evanescent modes. 
The penetration of evanescent modes is also discussed. 
The discussion is focused on two effects of ill-contact 
in electrodes, the evanescent modes and the nonlinear conductance.
The conclusion is given in {\S}~\ref{conclusion}. 

\section{Theory of Conductance in Nonequilibrium Green's Function} 
\label{SEC:THEORY}

We proposed a method of the eigen-channel decomposition of several properties 
of quantum transport in nanometer-scale conductors 
by using the nonequilibrium Green's function (NEGF) formalism.~\cite{caroli71, meir92}
The eigen-channel is the transmission eigenstate obtained 
by the unitary transformation of the original channels,~\cite{eigenchannel-landauer-buttiker} 
whose concept may be useful for analyzing current 
through nanometer-scale metal conductors. 
Brandbyge and coworkers obtained the expression for the eigen-channel decomposition 
of the local density of states (LDOS).~\cite{brandbyge98, eigenchannel-negf}
We generalize their formulation and provide the decomposition of NEGF 
and transport quantities, 
such as the density of electrons or local current, 
into contributions from respective eigen-channels. 
This formulation is also applied to the case of finite bias voltage.  
Information on the eigen-channel decomposition of local current  
may be necessary for analyzing the vortex of current.~\cite{nonoyama98}
Later in the present paper, we will clarify the contribution of 
\lq evanescent' modes by eigen-channel analysis. 

\subsection{Nonequilibrium Green's function}

We use a nanometer-scale metallic rod 
whose entire system consists of three parts, the conductor (C), 
the left lead wire (L), and the right lead wire (R). 
The two lead wires, L and R, are semi-infinite and connect with 
the central conductor C. 
The retarded and advanced Green's functions $G^{R}$ and $G^{A}$ 
in the conductor C can be obtained 
within the NEGF theory as~\cite{caroli71,meir92} 
\begin{eqnarray}
 G^{{R/A}}(E) 
= \left(E-H_{\mathrm{C}}-\Sigma^{{R/A}}_{\mathrm{L}}(E) 
                            - \Sigma^{R/A}_{\mathrm{R}}(E)\right)^{-1} \ ,
\label{eq:retarded-g}
\end{eqnarray}
where $H_{\mathrm{C}}$ is the Hamiltonian matrix of the conductor C  
and its matrix dimension $N_C$ is given 
as 
\begin{eqnarray}
 N_\mathrm{C} &=& {\rm (the\ number\ of\ atoms\ in\ the\ conductor) } \nonumber \\
&&\times {\rm (the\ number\ of\ orbitals\ per\ atom)}. 
\end{eqnarray} 
The coupling of C with the L(R) lead wires is represented by 
the retarded/advanced self-energies, $\Sigma^{R/A}_{\mathrm{L(R)}}$, 
which have non-zero elements only near the electrodes.
The lesser Green's function $G^{<}$ and 
the lesser self-energy $\Sigma^{<}$  are defined as ~\cite{caroli71,meir92} 
\begin{eqnarray}
 G^{<}(E) &=& G^{R}(E) \Sigma^{<}(E) G^{A}(E), 
\label{lesserG} \\
  \Sigma^{<}(E) &=& {\mathrm i} \left\{ f_{\mathrm{L}}(E) \Gamma_{\mathrm{L}}(E) +
 f_{\mathrm{R}}(E) \Gamma_{\mathrm{R}}(E) \right\},
\label{eq5}
\end{eqnarray}
where 
$\Gamma_{\mathrm{L (R)}}$ is a semi-positive-definite matrix given as 
$\Gamma_{\mathrm{L (R)}} = {\mathrm i} (\Sigma^{R}_{\mathrm{L (R)}} 
- {\Sigma^{R}_{\mathrm{L (R)}}}^{\dag})$ 
and $f_{\mathrm{L (R)}}$ is the Fermi-Dirac distribution function 
in the unperturbed lead wire L (R) of infinite size.
The lesser self-energy 
represents the rate of electron scattering by the electrodes.~\cite{Datt-book} 
We will calculate the self-energies  
by an efficient recursion method~\cite{lopez84} 
with an appropriate smearing factor $\delta=1.0\times10^{-5} {\rm Ry}$ and 
the temperature of electrons in the electrodes is assumed zero.
It should be noted that strictly localized states in C 
without interaction with incoming/outgoing modes in the lead wires 
are not included in $G^{<}$.~\cite{brandbyge02}

\subsection{Channel decomposition of physical properties into eigen-channels}

The channel-decomposition method was introduced 
for the local density of states (LDOS) by Brandbyge and coworkers.~\cite{brandbyge98, eigenchannel-negf} 
Now, we are generalizing the method by defining the channel decomposition of NEGF. 
The theory is formulated with real-space (atomic) bases. 
The transmission amplitude matrix from L to  R, $t_{\mathrm{RL}}$, 
and that from R to L, $t_{\mathrm{LR}}$, 
are given as~\cite{cuevas97}
\begin{eqnarray}
  t_{\mathrm{RL}}&\equiv& {\mathrm i} \Gamma_{\mathrm{R}}^{\frac{1}{2}}G^{R}_C\Gamma_{\mathrm{L}}^{\frac{1}{2}}, \label{eq:t_rl-negf}\\
  t_{\mathrm{LR}}&\equiv& {\mathrm i} \Gamma_{\mathrm{L}}^{\frac{1}{2}}G^{R}_C\Gamma_{\mathrm{R}}^{\frac{1}{2}}. \label{eq:t_lr-negf}
\end{eqnarray}
An incoming wavefunction of the eigen-channel ``$n$'' from the left electrode 
is denoted as the vector $u_{\mathrm{L}n}$, 
which is an eigen-vector of the matrix $t_{\mathrm{RL}}^{\dag} t_{\mathrm{RL}}$. 
The incoming wavefunction from the right electrode, $u_{\mathrm{R}n} $, 
is defined in the same manner. 
We can define the decomposition of the lesser self-energy $\Sigma^{<}$ 
into contributions  from respective eigen-channels 
$u_{\mathrm{L}n}$ and $u_{\mathrm{R}n}$; 
\begin{eqnarray}
	\Sigma^{<} \nonumber &=& {\mathrm i} f_{\mathrm{L}} \Gamma_{\mathrm{L}} + {\mathrm i} f_{\mathrm{R}} \Gamma_{\mathrm{R}} \nonumber\\
	&=& \sum_n({\mathrm i} f_{\mathrm{L}} \Gamma_{\mathrm{L}}^{\frac{1}{2}}u_{\mathrm{L}n}u_{\mathrm{L}n}^\dag \Gamma_{\mathrm{L}}^{\frac{1}{2}} + {\mathrm i} f_{\mathrm{R}} \Gamma_{\mathrm{R}}^{\frac{1}{2}}u_{\mathrm{R}n}u_{\mathrm{R}n}^\dag \Gamma_{\mathrm{R}}^{\frac{1}{2}})\nonumber\\
	&=& \sum_n \Sigma^{<}_{\mathrm{L}n} + \sum_n \Sigma^{<}_{\mathrm{R}n}.
\label{eq:decomposed-lesser-se-sum} 
\end{eqnarray}
This definition is a basic idea and, then, 
the decomposition of the lesser Green's function $G^{<}$ 
can be introduced as 
\begin{eqnarray}
	G^{<} &=&G^R \Sigma^{<} G^A 
	= \sum_n G^{<}_{\mathrm{L}n} + \sum_n G^{<}_{\mathrm{R}n},  
\label{eq:decomposed-lesser-g}
\end{eqnarray}
where $G^{<}_{\mathrm{L(R)}n}$ is defined as
\begin{eqnarray}
 G^{<}_{\mathrm{L}n} =  G^R \Sigma^{<}_{\mathrm{L}n}G^A , \ \ \ 
 G^{<}_{\mathrm{R}n} =  G^R \Sigma^{<}_{\mathrm{R}n}G^A . 
\label{eq:decomposed-lesser-g2}
\end{eqnarray}

By using this eigen-channel decomposition, 
the  local electron density $\rho_i(E)$~\cite{caroli71,meir92,lopez84} 
is decomposed into eigen-channel components;  
\begin{eqnarray}
 \rho_i(E) 
   &=& \frac{1}{2\pi}{\rm Im} G^{<}(i ,i)(E) \nonumber \\
   &=& \sum_n \rho_{\mathrm{L}n}(i, E) + \sum_n \rho_{\mathrm{R}n}(i, E)  ,
\label{eq:epdens}
\end{eqnarray}
where $\rho_{\mathrm{L}n}(i, E)$ is defined as
\begin{eqnarray}
\rho_{\mathrm{L}n}(i, E) &\equiv& 
\frac{1}{2\pi}{\rm Im} G^{<}_{\mathrm{L}n}(i,i)(E) \nonumber\\
&=& \frac{1}{2\pi} f_{\mathrm{L}} {\rm Re} [ 
(G^R \Gamma_{\mathrm{L}}^{\frac{1}{2}}u_{\mathrm{L}n}
u_{\mathrm{L}n}^\dag \Gamma_{\mathrm{L}}^{\frac{1}{2}} G^A )_{ii} ] \nonumber\\
&=& \frac{1}{2\pi} f_{\mathrm{L}} {\rm Re} [ 
(G^A \Gamma_{\mathrm{L}}^{\frac{1}{2}}u_{\mathrm{L}n})_i
(G^R \Gamma_{\mathrm{L}}^{\frac{1}{2}}u_{\mathrm{L}n})_i ]
\end{eqnarray}
and $\rho_{\mathrm{R}n}(i, E)$ is defined in the same manner. 
The local current $i_{i \leftarrow j}(E)$~\cite{caroli71,meir92,lopez84} 
is also decomposed, 
when the energy $E$ lies within the chemical potentials 
at the left and right electrodes ($\mu_{\mathrm{R}} < E < \mu_{\mathrm{L}}$);
\begin{eqnarray}
i_{i \leftarrow j}(E) 
&=& 2 \frac{G_0}{\mathrm e} {\rm Re} [H_\mathrm{C}(i, j)G^{<}_{j, i}(E)]  \nonumber \\
&=&\frac{G_0}{\mathrm e}\sum_n i_{n, i \leftarrow j}(E),
\label{def:l-curr-dec}
\end{eqnarray}
where
\begin{eqnarray}
i_{n, i \leftarrow j}(E) &\equiv& 
2{\rm Re} [ H_{j i} G_{n}^{<}(i, j)(E)]  \nonumber \\
&=& -2 {\rm Im} [H_{j i} 
(G^R \Gamma_{\mathrm{L}}^{\frac{1}{2}}u_{\mathrm{L}n})_i
(G^R \Gamma_{\mathrm{L}}^{\frac{1}{2}}u_{\mathrm{L}n})^*_j]. \nonumber \\
\label{def:l-curr}
\end{eqnarray}
Here, 
$G_0$ is the quantum unit of conductance including spin degeneracy 
($G_0=\frac{2e^2}{h}\simeq 0.775 \times 10^{-4}~ \Omega^{-1}$) 
for spin-unpolarized calculations.~\cite{caroli71, meir92, nonoyama98}
The eigen-channel decomposition of LDOS by Brandbyge et al.~\cite{eigenchannel-negf} 
is given by substituting $f_\mathrm{L}=f_\mathrm{R}=1$ 
into the expression of the local electron density, eq.~(\ref{eq:epdens}). 

\section{Evanescent Wave Modes and Length Dependent Conductance}
\label{SEC:TOTAL-DECAY-BEHAVIOR}

The presence of the evanescent wave current and nonlinear conductance 
is demonstrated, in this section, 
in a nanometer-scale conductor with an electronic structure 
of s and/or d orbitals, 
which may be a prototypical model for realistic metals.
Here, we adopt the Hamiltonian of spin-unpolarized bulk (fcc) gold (Au) 
only with s-s or d-d hopping integrals without s-d hybridization 
in the framework of the tight-binding linear muffin-tin orbital 
(LMTO) method with the atomic sphere approximation 
(ASA),~\cite{andersen75,andersen84,HighligtLMTO}  
specifically the first-order Hamiltonian $H^{(1)}$. 

\subsection{Gold nanorods: calculated systems}

\begin{figure}[t]
\centering
\includegraphics[width=8.8cm]{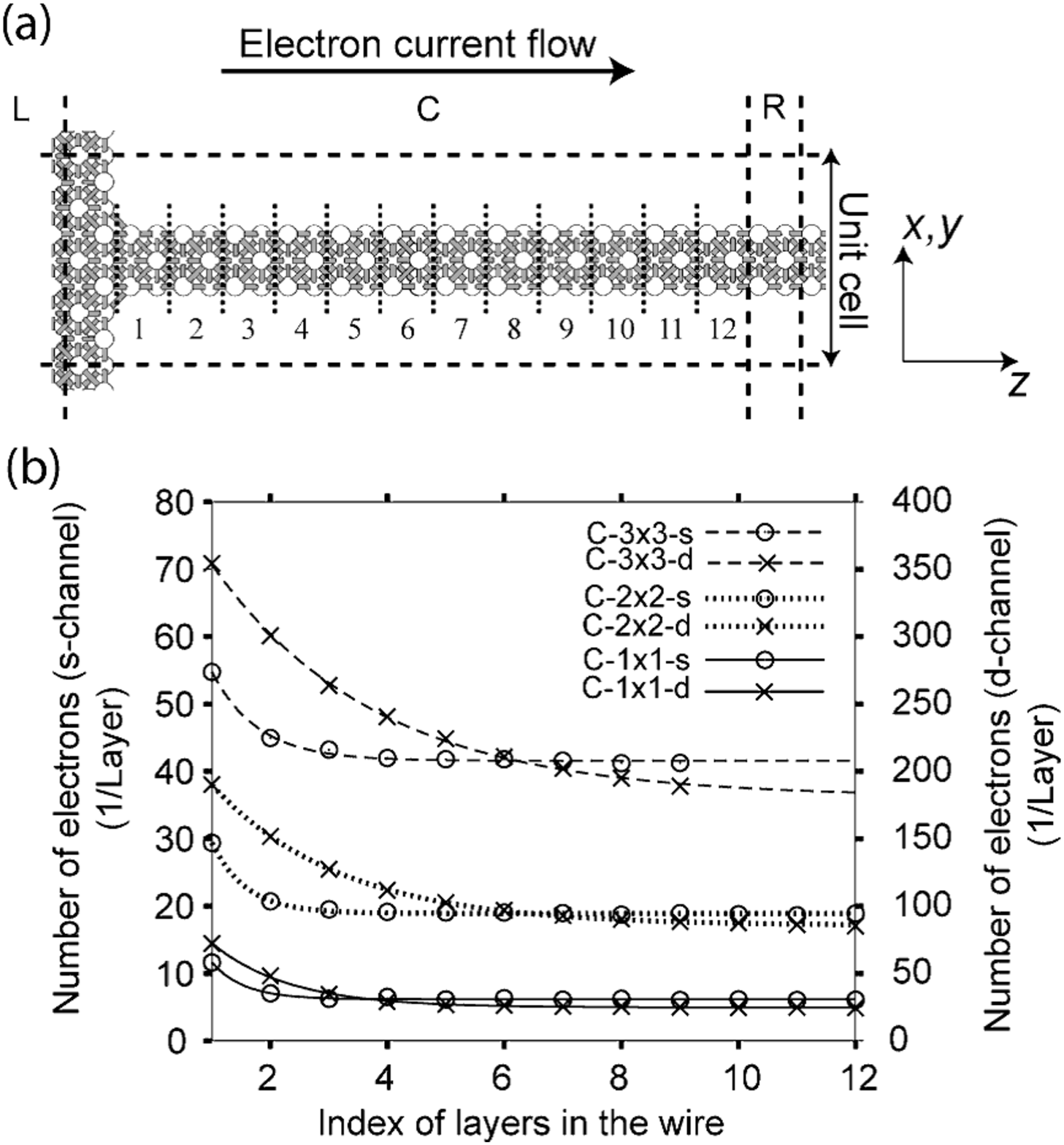}
\caption{
(a) Calculated system of a C-$1 \times 1$ structure. 
The conductor region (C) has
$k=1,2,3....12$ periodic layers. 
Here, a periodic layer  is defined as two atomic layers 
(see text for details)
(b) Decaying behavior of  the averaged current mode,  
due to the presence of evanescent wave components. 
The number of electrons per periodic layer
$\rho^{(k)}$, 
defined in eq.~(\ref{EQ-RHO-LAYER}), is plotted 
as a function of the number of periodic layers. 
No bias voltage is applied, and 
electrons enter the system only from the left electrode
($f_\mathrm{L}=1$, $f_\mathrm{R}=0$). 
\label{fig:decay2}
}
\end{figure}

\begin{figure}[t]
\centering
\includegraphics[width=.475\textwidth,clip]{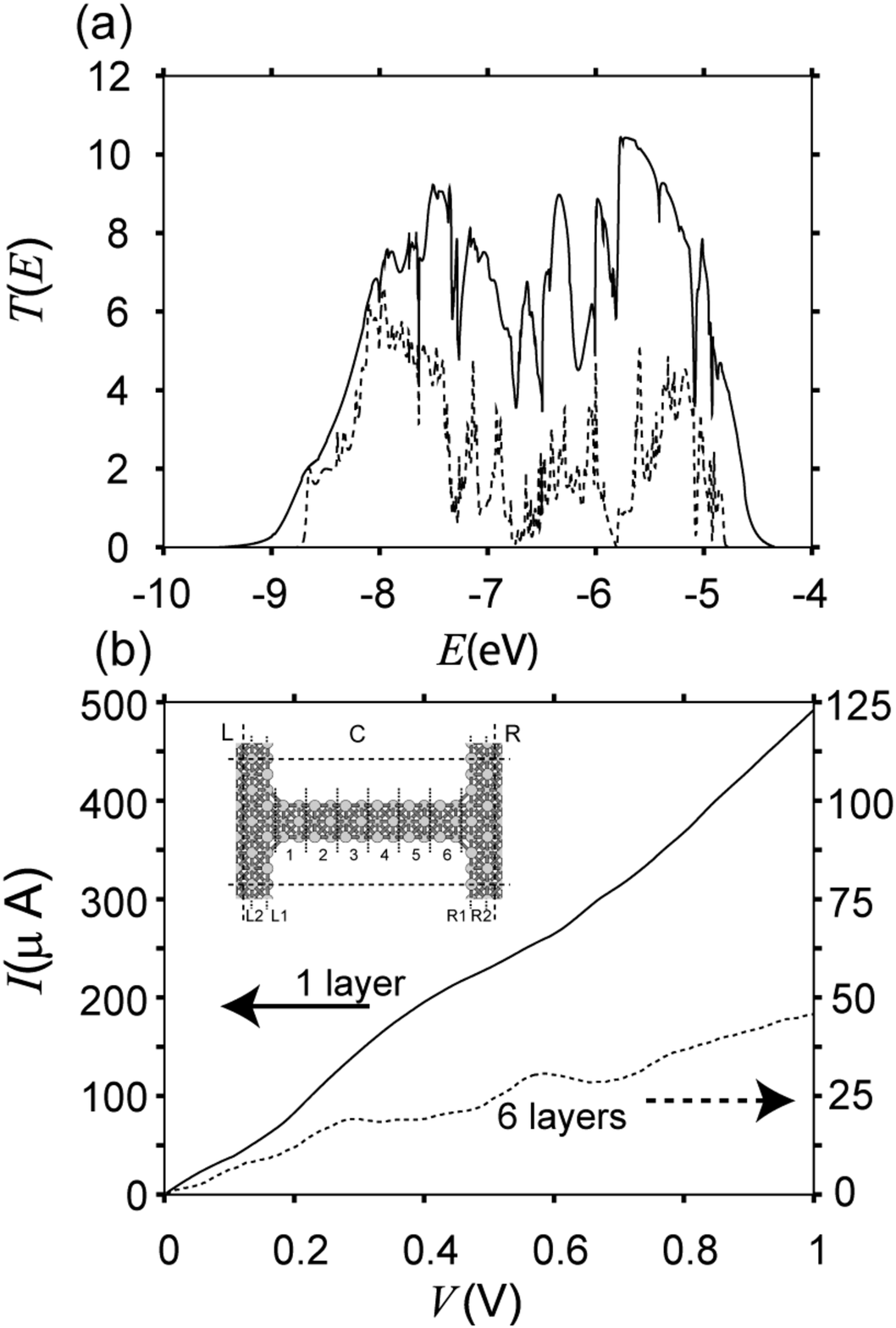}
\caption{Length dependence of transport properties for the d-channel system 
of C-$1\times 1$ cross section: 
(a) Transmissions $T(E)$ of the six-layer conductor (the broken line) and 
the 1-layer conductor (the solid line) 
at $V=0~{\rm V}$, both assuming $\mu_{\rm R}=-6.6$ eV. 
The transmission of the shorter rod is larger, owing to the contribution 
by evanescent wave modes.
(b) $I$-$V$ curves of the six-layer conductor (the broken line) and 
the two-layer conductor (the solid line). 
Here, we assume $\mu_{\rm R}=-6.6$ eV.
The longer rod shows a higher nonlinearity because of the enhanced spiky structure 
of the transmission function.
The inset shows the geometry of the six-layer conductor. 
\label{fig:struc3x2-s-etrans2D}
}
\end{figure}

Calculated systems are nanometer-scale conductors: 
(1) systems of semi-infinite size extending to $z=+\infty$  
connecting with a left lead wire, as depicted in Fig.~\ref{fig:decay2}(a)
and 
(2) those of finite size connecting with left and right lead wires, 
as shown in the inset in Fig.~\ref{fig:struc3x2-s-etrans2D}(b). 
The conductors and lead wires are quarried 
from an ideal gold fcc lattice with a lattice constant of $a=7.755$~a.u.  
Actual gold nanowires have helical multishell nanostructures~\cite{TAKAYANAGI2000}
whose formation process is explained with the two-stage process 
based on the quantum molecular dynamical simulation.~\cite{IGUCHI2007} 
However, in the present paper, we use the above toy model of gold nanorods 
with an ideal fcc structure, without lattice relaxation nor reconstruction. 
The left lead wire L consists of two layers of 64 atoms and 
the right lead wire R is the conductor itself of semi-infinite size 
(Fig.~\ref{fig:decay2}(a)) or  
the left and right lead wires L and R are 
the same two-layer blocks of 64 atoms 
of finite size (Fig.~\ref{fig:struc3x2-s-etrans2D}(b)). 
The cross sections of the wires C are  $1 \times 1$, $2 \times 2$, and 
$3 \times 3$ fcc conventional unit cells. 
We denote the structures by C-$1 \times 1$, C-$2 \times 2$, and C-$3 \times 3$, 
respectively. 
The lengths of the wires are 12 layers for C-$1\times 1$ and C-$2\times 2$, 
and 9 layers for C-$3\times 3$.   
Each layer consists of two sublayers because of the fcc structure, 
one atom layer contains five atoms and the other one contains four atoms 
in each unit ($1 \times 1$). 

The periodic boundary condition is imposed 
on the $x-y$ plane with the primitive vectors 
($4a$, $0$, $0$) and ($0$, $4a$, $0$), 
and only the $\Gamma$ point in the $k_x$-$k_y$ plane is used 
for the electronic structure calculations. 
In other words, the cross section of the lead wires is of periodic $4 \times 4$. 
Effects of a bias voltage $V$ are added to the Hamiltonian.  
In the present work, the LDA potential is not determined self-consistently 
in conductors at a finite bias voltage and 
a uniform voltage drop of a bias potential is assumed within a conductor. 
The voltage drop in a conductor may not be uniform in actual nanometer-scale materials 
but may occur near the regions of contacts 
where charge distribution would change largely,~\cite{eigenchannel-negf,tsukamoto02} 
and the self-consistent calculation for LDA potential in thick nanometer-scale wires 
may be important for conducting detailed quantitative analysis. 
The structure of voltage drop presumably depends on the conditions 
of electrodes and the conductor; for example, in ill-connected electrodes, 
the voltage drop occurs in the region of electrodes. 

\subsection{Decaying electron density in rods and evanescent mode}

We show the electron density for s- and d-channels 
in Fig.~\ref{fig:decay2}(b) for several semi-infinite size rods 
(Fig.~\ref{fig:decay2}(a)) of different cross sections, 
in which the local electron density ($\rho_j(E)$) is integrated 
with respect to energy and 
plotted as the sum of each periodic layer; 
\begin{eqnarray}
\rho^{(k)} = \int_{-\infty}^{+\infty}
\sum_{j\in {k {\rm -th} \, {\rm layer}}} \rho_j(E) {\mathrm d} E    , 
\label{EQ-RHO-LAYER}
\end{eqnarray}
and we observe the decaying characteristics of electron density.  
Here, no bias voltage is applied, and 
electrons enter the system only from the left electrode 
($f_\mathrm{L}=1$, $f_\mathrm{R}=0$). 
The quantity of $\rho^{(k)}$ shows 
the averaged characteristic among all existing modes. 
We also calculated the electron density profile in similar periodic systems 
with a larger cross section of the lead wires of $6\times 6$, 
and no appreciable change was observed. 
Therefore, we can conclude that the above dependence 
on the cross section of conductors is universal but not 
a size effect of the adopted systems.  
The decaying characteristics of the electron density can be expressed 
as 
\begin{eqnarray}
\rho^{(k)} = \rho^{(\infty)} + \rho^{(0)}{\rm e}^{-\frac{k}{\xi}}  ,
\label{def:xi}
\end{eqnarray}
and we can define the penetration length $\xi$. 
Figure \ref{fig:decay2}(b) indicates the existence of 
components of evanescent wavefunction among the modes, 
which are near-field standing wavefunctions 
decaying exponentially with increasing distance from the electrode. 
The components of evanescent wave modes decay owing to backscattering     
and, therefore, 
only Bloch wave modes reach the right end ($z=+\infty$). 
In other words, 
when the conductor length is comparable to or smaller than the decay length, 
the evanescent wavefunctions can reach the opposite electrode.
Therefore, one can expect the length dependence on  conductance. 
Note that the comparison between 
the s- and d-orbital channels is given later 
in {\S}~\ref{SEC:DISCUSSION}.

\subsection{Nonlinear conductance with finite bias voltage}

The transmission function $T(E)$ is defined as~\cite{caroli71,meir92,lopez84} 
\begin{eqnarray}
T(E) = {\rm tr} \big[ G^A(E)\Gamma_{\rm R}(E)G^R(E)\Gamma_{\rm L}(E)\big]  .
\end{eqnarray}
By using eqs.~(\ref{eq:t_rl-negf}) and (\ref{eq:t_lr-negf}), 
the transmission function $T(E)$ can be decomposed 
into the eigen-transmission 
$|\tau_n|^2$ with the unitary matrix $U_\mathrm{L}$ 
in terms of diagonalization of 
$t_{\mathrm{RL}}^{\dag} t_{\mathrm{RL}}$,~\cite{martin92}
\begin{eqnarray}
 T(E) &=& {\rm tr}\left[ t_{\mathrm{RL}}^{\dag} t_{\mathrm{RL}}\right]\nonumber = \sum_n |\tau_n|^2,\\
 U_{\mathrm{L}}^\dag t_{\mathrm{RL}}^{\dag} t_{\mathrm{RL}} U_{\mathrm{L}} &=& {\rm diag} \{|\tau_n|^2\}\label{eq:diag}. 
\end{eqnarray}
The incoming wavefunctions of the eigen-channel from L (the left electrode) can be given 
by the components of the unitary matrix $U_\mathrm{L}$ or $u_{\mathrm{Ln}}$.
The current can be evaluated by using the transmission function 
with the relation~\cite{caroli71,meir92,lopez84} 
\begin{eqnarray}
I(V) &=& \frac{G_0}{\mathrm e}\int_{\mu_{\mathrm{R}}}^{\mu_{\mathrm{R}}+ {\mathrm e} V} T(E) {\mathrm d} E , 
\end{eqnarray}
where $V$ is the bias voltage applied to the system, 
$\mu_{\mathrm{L (R)}}$ is the chemical potential of the L  (R) lead wire 
($\mu_{\mathrm{L}} = \mu_{\mathrm{R}} + eV, V > 0$).
The length dependence in the conductance of a d-orbital channel 
is shown in Figs.~\ref{fig:struc3x2-s-etrans2D}(a) and \ref{fig:struc3x2-s-etrans2D}(b) 
for systems of different rod lengths with a C-$1\times 1$ cross section. 
The calculated conductance in Fig.~\ref{fig:struc3x2-s-etrans2D}(a) shows 
that the shorter rod has a larger conductance 
owing to the presence of the evanescent wave modes.
In longer systems, the Bloch modes only contribute to conductance. 
Here, we notice a larger nonlinearity in model systems in the $I$-$V$ curve of 
Fig.~\ref{fig:struc3x2-s-etrans2D}(b), 
because  a  strong resonance scattering in the d-orbital channel 
causes a strong energy dependence 
in the transmission  function $T(E)$, which will be seen in the next section.

\section{Origin of Nonlinear Conductance and Evanescent Modes}
\label{origin}
\subsection{Quantization of one-dimensional modes along rod length 
in d-orbital channels}\label{n-con}

Now, we investigate characteristics of the eigen-channels 
by generalized channel analysis in {\S}~\ref{SEC:THEORY}.

\begin{figure}[t]
 \centering
 \includegraphics[width=.45\textwidth,clip]{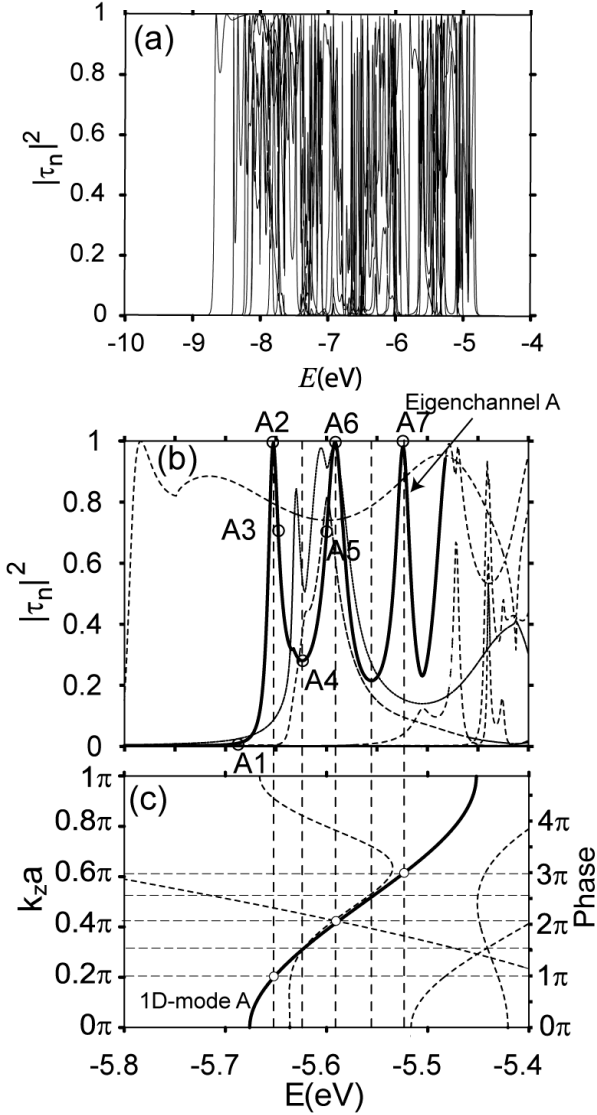}
 \caption{Eigen-transmission for the d-channel system in a narrow energy range. 
(a) Eigen-transmission of the six-layer conductor at $V=0~{\rm V}$ and 
(b) eigen-transmission as a function of energy near the d-band bottom.  
 The eigen-channel A is shown by a thick line. 
 (c) $E$-$k$ curve of an infinite conductor and the phase of 
one-dimensional modes (right-hand side) in the conductor, 
which is calculated as $N_{\rm eff}a \times k_z$ ($N_{\rm eff}=4.88$), 
where $N_{\rm eff}a$ is the effective length of the conductor. 
The open circles in (c) are the points used in the fitting procedure of $N_{\rm eff}$ 
and correspond to A2, A6, and A7 in (b). 
 }
\label{fig:struc3x12-d-etrans2D-2}
\end{figure}   

Figure \ref{fig:struc3x12-d-etrans2D-2}(a) 
shows the eigen-transmissions $|\tau_n|^2$ 
of the six-layer conductor in Fig.~\ref{fig:struc3x2-s-etrans2D}, 
which shows a strong energy dependence or a spiky structure 
of an energy interval of $0.01 \sim 0.02$~eV.
The eigen-transmission within a narrow energy range 
($-5.8~{\rm eV}<E<-5.4~{\rm eV}$) 
is depicted in Fig.~\ref{fig:struc3x12-d-etrans2D-2}(b), 
in which we focus on the eigen-channel \lq A'. 
The oscillatory behavior of the transmission function of the eigen-channel A 
is due to the resonance and off-resonance of one-dimensional Bloch modes. 
The d-bands in bulk systems have a width of about 4~eV. 
Then, the finiteness of the rod radius in the $x$-$y$ plane makes each band split. 
This splitting causes a complicated band-crossing since some d-bands exhibit the 
dispersion relation ${\rm d}E/{\rm d}k_z>0$, whereas 
others exhibit the dispersion relation ${\rm d}E/{\rm d}k_z<0$. 
Then, a complicated {\it branching} occurs in d-orbital systems. 
We plot, in Fig.~\ref{fig:struc3x12-d-etrans2D-2}(c), 
the $E$-$k_z$ curve of the $1 \times 1$ infinite conductor 
representing the {\it branch} of a one-dimensional mode  \lq A' 
corresponding to the `A'-channel. 

\begin{figure}[t]
 \centering
 \includegraphics[width=.475\textwidth,clip]{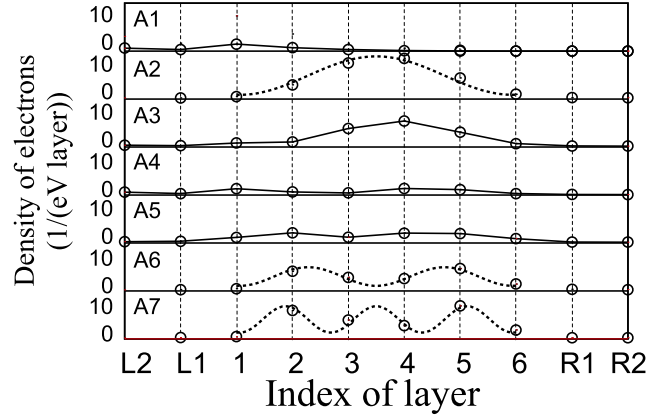}
 \caption{Eigen-channel decomposed 
 density of electrons for the eigen-channels 
 from A1 to A7 in Fig.~\ref{fig:struc3x12-d-etrans2D-2}(b). 
 Electrons enter the system only from the left lead wire 
 ($f_\mathrm{L}=1$, $f_\mathrm{R}=0$). 
Broken lines show the function $a\sin^2(\frac{n\pi}{N_{\rm eff}}(x-3.5+\frac{N_{\rm eff}}{2}))+b$ 
with the layer index $x$ and $n=1,2,$ and 3 for A2, A6, and A7. 
The adjustable parameters (a,b) are (8.0, 0.9), (3.9, 1.1), and (5.6, 1.2) 
for A2, A6, and A7, respectively. 
These lines suggest that the charge densities 
at A2, A6, and A7 may have one, two, and three peaks, respectively, 
as the phase in Fig.~\ref{fig:struc3x12-d-etrans2D-2}(c) shows.
 }
  \label{fig:channel-wave}
\end{figure}

The right vertical axis refers to the phase $N_{\rm eff}a\times k_z$ of 
the branch `A' in an infinite periodic wire 
of fitted effective periodicity length $N_{\rm eff}=4.88$. 
In fact, the periodic unit length $N_{\rm eff}a$ of 4.88 is determined 
so that the perfect transmission ($|\tau_n|^2=1$) 
occurs at the phases of $\delta = \pi,\ 2\pi$, and 3$\pi$.  
This $N_{\rm eff}$ value (4.88) is nearly identical to 
that of the conductor part ($N=6$). 
The fitted length unit of $N_{\rm eff}=4.88$ is actually shorter than 
that of the actual conductor part ($N=6$), 
since the wavefunctions are scattered in the regions near the electrode 
and the resultant effective conductor length 
should be shortened from that of the ideal one. 
The above analysis indicates that 
each eigen-channel ({\it branch}) of the d-orbital system has peaks of the transmission 
corresponding to a phase 
at $N_{\rm eff}a\times k_z=n\pi$. 
The energy interval between {\it branches} is determined by 
the size of the rod and the structure of d-bands and 
should be inversely proportional to the radius of the rod. 
In other words, we have observed two types of nanometer-scale rod quantization; 
the quantization of the radius of the rod ({\it branch}) and 
that of the length of the rod (peak structure of eigen-transmission). 
The oscillatory behavior of resonance and off-resonance in s-orbital channels 
is relatively weak;~\cite{lee04} 
one branch extents over a much larger energy range. 
For example, the energy width of one branch is $2\sim 4$~eV and 
the energy width of the s-band is about 10~eV in Au.  
The energy interval of transmission peaks is wider, 
and the difference between the maximum and minimum values is small. 
This oscillatory behavior difference may be 
consistent with the decaying profile of electron density 
in s- and d-orbital channels in Fig.~\ref{fig:decay2}(a), 
{\it i.e.}, 
the evanescent wavefunctions appear in order to connect with wavefunctions 
in the conductor and lead wires in the electrode region. 
In the d-orbital channels, more evanescent wavefunctions 
are necessary than in s-orbital channels, 
because the wavefunctions have more freedoms of orbital. 
The strong resonance scatterings causing the large oscillatory behavior 
in d-orbital channels are consistent with previous works 
on transition metals.~\cite{Tekman1989,lee04,pauly06}
The evanescent wavefunctions of d-orbital channels 
penetrate much deeper and are more scattered at the electrodes, 
in order to connect with Bloch modes. 
This we call the `ill-contact' of d-orbital channels. 
The channel decomposition of the local density of electrons 
is shown in Fig.~\ref{fig:channel-wave}
for several energy points A1$\sim$A7 of the eigen-channel `A'  
in Fig.~\ref{fig:struc3x12-d-etrans2D-2}(b). 
We can see the characteristics consistent with the above mode analysis results; 
e.g.,  
the point A1 is placed at the tail position 
of the branch `A' in Fig.~\ref{fig:struc3x12-d-etrans2D-2} (b) 
and its local density of electrons 
exhibits a typical decaying (evanescent) behavior.  
The point A2 is placed at a peak of the eigen-transmission 
of the branch `A' in Fig.~\ref{fig:struc3x12-d-etrans2D-2}(b) 
that exhibits a nodeless standing wave, 
which is consistent with the assignment of the phase of $\delta=\pi$ 
in Fig.~\ref{fig:struc3x12-d-etrans2D-2}(c). 
The density profile shows that the connection of transmission 
in C and R or L is poor and does not smear out from C.

\begin{figure}
\centering
\includegraphics[width=.45\textwidth,clip]{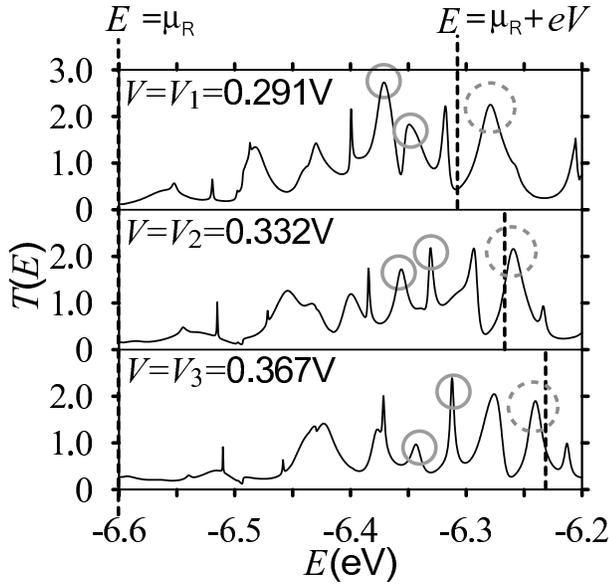}
\caption{Voltage dependence of transmission function $T(E)$ for three cases: 
$V_1=0.291$~V, $V_2=0.332$~V, and $V_3=0.367$~V. 
The chemical potential $\mu_{\rm R}=-6.6$~eV 
and respective energies $\mu_{\rm R}+{\mathrm e} V$ are shown by broken lines.
The transmission function $T(E)$ between  $\mu_{\rm R}$ and $\mu_{\rm L}=\mu_{\rm R}+{\mathrm e} V$ 
contributes to the current for the bias voltage $V$. 
The peak shown by a broken circle creates the structure of 
$\left(\frac{{\mathrm d} I}{{\mathrm d} V}\right)_T$ in the $I$-$V$ curve. 
The peak structures marked by solid circles contribute strongly 
to $\left(\frac{{\mathrm d} I}{{\mathrm d} V}\right)_{\Delta T}$.}
\label{fig:struc3x12-d-trans2}
\end{figure}

\subsection{Nonlinear conductance and bias voltage dependence of transmission}
\label{SEC:DISCUSSION0}

Let us discuss the finite voltage case. 
The transmission function $T(E)$  strongly depends on the bias voltage $V$, as shown 
in Fig.~\ref{fig:struc3x12-d-trans2}. 
Furthermore, there exist two characteristic features of d-orbital channels: 
(i) the negative differential conductance ($\frac{{\mathrm d} I}{{\mathrm d} V}<0$) and 
(ii) the peak structure in the differential conductance.
The differential conductance is separated into two parts:
\begin{eqnarray}
 \frac{{\mathrm d} I(V)}{{\mathrm d} V} &=& G_0 T(E=\mu_\mathrm{R}+{\mathrm e} V, V) \nonumber \\
 &&+\frac{G_0}{e} \int_{\mu_\mathrm{R}}^{\mu_\mathrm{R}+eV}\frac{\partial T(E,V)}{\partial V}{\mathrm d} E  
\nonumber\\
 &=& \left(\frac{{\mathrm d} I}{{\mathrm d} V}\right)_T + \left(\frac{{\mathrm d} I}{{\mathrm d} V}\right)_{\Delta T} .
\label{eq:diff-i}
\end{eqnarray}
The first term  $\left(\frac{{\mathrm d} I}{{\mathrm d} V}\right)_T$ 
is the contribution from a new structure of the transmission function 
$T(E)$ entering  the energy window ($\mu_{\rm R}+{\mathrm e} V>E>\mu_{\rm R}$) 
from the higher energy side and is always positive. 
The second term  $\left(\frac{{\mathrm d} I}{{\mathrm d} V}\right)_{\Delta T}$ 
is the contribution from the change of the overall structure of the transmission 
function $T(E)$ in the energy window and can be negative.
Therefore, the negative differential conductance is 
due to the strong voltage dependence of the overall structure of the transmission function, 
which is caused by the channels confining the local electron density 
within the conductor. 
The sensitivity of the resonance to the bias voltage is 
due to the small width of branches of d-bands as 
a result of ill-contact of d-orbital channels. 
The peak entering the energy window 
of $\mu_R+{\mathrm e} V > E > \mu_R$ from the higher energy side  
gives rise to the peak of $\frac{{\mathrm d} I}{{\mathrm d} V}$. 
Therefore, the peak structure of differential conductance originates 
from that of the transmission due to the resonance scattering.

\subsection{Quantization due to finiteness of radius of rod}
\label{SEC:DISCUSSION}

The mode analysis in {\S}~\ref{n-con} 
indicates that 
the decaying behavior in Fig.~\ref{fig:decay2}(b) appears 
not only in specific decay modes but also in the sum of various modes,
such as those in Fig.~\ref{fig:channel-wave}.

\begin{figure}[t]
\centering
\includegraphics[width=.45\textwidth,clip]{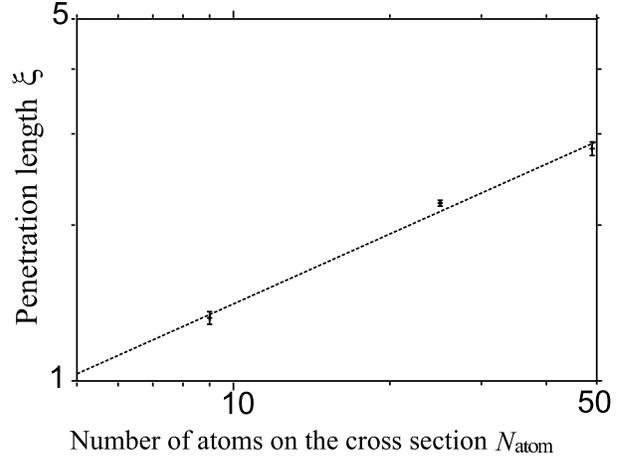}
\caption{
Dependence of penetration lengths $\xi$ 
on the number of atoms on cross section of the conductor $N_{\rm atom}$ 
for d-orbital channels for the local electron density $\rho^{(k)}$ 
in Fig.~\ref{fig:decay2}(b). 
The points in the figure indicate the values calculated using eq.~(\ref{eq:L-dep}) 
with error bars 
and the dotted line shows the behavior of $N_{\rm atom}^{0.45\pm 0.04}$ 
determined by the least-square fitting procedure. 
}
\label{fig:L-dep}
\end{figure}

Now, we explain the physical picture of the difference between s- and d-channels 
for the decaying behavior in Fig.~\ref{fig:decay2}. 
Here, we denote the number of bases of electron wavefunctions 
on the cross section of the conductor $N_{\rm cross}$,  
the energy width of the s- or d-band $W_{\rm band}$ and 
that of one branch $W_{\rm branch}$. 
For example, one can see from Fig.~\ref{fig:struc3x12-d-etrans2D-2} that
$W_{\rm band} \simeq 4$~eV (the width of the d-bands) 
and $W_{\rm branch} \simeq 0.2$~eV (the {\it branch} A). 
$N_{\rm cross}$ should be proportional to the number of atoms 
on the cross section of the conductor $N_{\rm atom}$. 
A mode can contribute to the current, 
when it connects with a Bloch wavefunction in the conductor C in the contact region. 
The number of one-dimensional bands in the conductor C 
can be estimated as 
\begin{eqnarray}
N_{\rm cross}\frac{W_{\rm branch}}{W_{\rm band}}. 
\label{EQ-MODE-NUMBER}
\end{eqnarray}
If the ratio  
\begin{eqnarray}
R \equiv  \frac{W_{\rm branch}}{W_{\rm band}} 
\label{EQ-MODE-NUMBER2}
\end{eqnarray}
is small, 
it is difficult for incoming modes to connect with the one-dimensional Bloch modes 
in C and 
majority of the incoming waves may be backscattered.  
The ratio $R$ is, in our examples,  
of the order of $0.2/4 \simeq 0.05$ in the d-orbital channels.   
The above explanation is consistent with the data in Fig.~\ref{fig:decay2},
since the contribution of evanescent modes can be estimated 
as  the difference in local electron density 
between the left end ($z=1$) and the right end ($z=\infty$).
In the case of the C-$3 \times 3$-d system,
for example, 
the numbers of electrons at the left and right ends
are $n_{\rm left} \approx 350$ and 
$n_{\rm right} \approx 180$, respectively. 
The contributions of (one-dimensional) Bloch modes ($n_{\rm Bloch} \approx n_{\rm right}$) 
and evanescent modes ($n_{\rm Evan} \approx n_{\rm left} - n_{\rm right}$)
are almost the same. 
The evanescent modes are very important for the d-orbital channels, 
since $n_{\rm Evan} / n_{\rm Bloch} \approx 1$. 
The same analysis of the s-band system indicates that 
the evanescent modes are less important ($n_{\rm Evan} / n_{\rm Bloch} < 1 $) 
than that in the d-orbital channels,
which is consistent with the fact that 
the ratio 
$R  = W_{\rm branch} / W_{\rm band}$ is larger than that in the d-orbital channels 
as $R\simeq 0.2 \sim 0.4$ 
because $W_{\rm band}\simeq 10~{\rm eV}$ and  
$W_{\rm branch}\simeq 2\sim 4~{\rm eV}$.  
The discussion of the connection is supported 
by the dependence of penetration length 
on the number of atoms in the cross section.
For simplicity, an extreme case with dispersionless sub-bands 
is considered for d-orbital channels. 
In this case, the mean energy interval $E_{\rm int}$ at Fermi energy is scaled by \
\begin{eqnarray}
E_{\rm int} \propto \frac{1}{N_{\rm cross}}.
\end{eqnarray} 
Because the penetration length $\xi$ can be estimated as
\begin{equation}
\xi \propto \frac{1}{\sqrt{E_{\rm int}}},
\end{equation} 
we obtain 
\begin{equation}
\xi\propto \sqrt{N_{\rm cross}} \propto \sqrt{N_{\rm atom}} \ \ .
\label{eq:L-dep}
\end{equation} 
Figure~\ref{fig:L-dep} shows the dependence of the penetration length $\xi$ 
on the number of atoms in the cross section $N_{\rm atom}$ 
for d-orbital channels 
for the local electron density $\rho^{(k)}$ in Fig.~\ref{fig:decay2}(b) 
with eq.~(\ref{def:xi}). 
Because the penetration length is almost scaled by eq.~(\ref{eq:L-dep}),
we can conclude that our simple explanation of the connection captures 
an essential nature of evanescent waves. 
It should be noted that a discussion of scattered waves 
gives the same dependence of the penetration length. 
Given incident waves are backscattered at edges of the conductor, 
the collision rate per periodic layer is proportional to ${1}/{L}$, 
where $L$ is the radius of the conductor. 
Because the penetration length is proportional to the inverse of the collision rate and 
$N_{\rm cross}\propto L^2$, eq.~(\ref{eq:L-dep}) is obtained again. 
This supports our discussion in this subsection. 

\begin{figure}[t]
 \centering
 \includegraphics[width=.45\textwidth,clip]{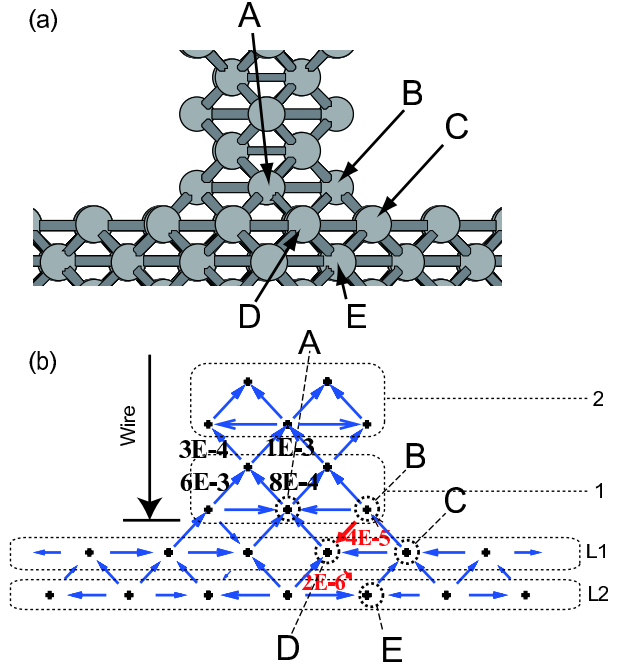}
 \caption{Channel-decomposed local current at the energy point A1 
 near the electrode region between the conductor and the lead wire;
 (a) Geometrical structure near the electrode. 
 (b) The channel-decomposed local currents are 
 plotted as arrows and the numbers indicate their values. 
 The lengths of arrows indicate their amplitudes on the logarithmic scale. 
 The backward current flows from atom D to atom E ($2\times 10^{-6}$) and 
 from atom B to atom D ($4\times 10^{-5}$) are several orders of magnitude weaker 
 than the forward flows. 
 } \label{fig:local_current}
\end{figure}

\subsection{Eigen-channel decomposition of local current and 
evanescent wave modes}
The backward flow cannot be observed 
even if we would try to observe the total current flow 
at a certain layer position  because of the conservation law of current.  
In other words, the net flow conserves at each atomic position. 
Furthermore, no backward flow is observed in the total current flow 
because of strong forward components of the other modes. 
Therefore, it is essential to decompose the local current 
into the eigen-channel components and, then, 
the eigen-channel current can show the backflow related to the evanescent modes.  
We show that the local current defined by eq.~(\ref{def:l-curr-dec}), 
using the generalized channel analysis, gives the backward current flow. 
Figure \ref{fig:local_current} shows the 
channel-decomposed local current in eq.~(\ref{def:l-curr}) for the channel A1 
in Fig.~\ref{fig:struc3x12-d-etrans2D-2}. 
The backward current flow is found,
for example, between the atoms $B$ and $D$
and between the atoms $D$ and $E$. 
The same analysis for the channels A2 and A3 
(figure not shown) indicates that 
the backward current flow appears commonly 
between the atoms $D$ and $E$
among the above three channels.
The above result supports the discussion in the previous section,
in which the evanescent wave component should be attributed 
not only to decay channels, such as the channel A1,  
but also to many other channels. 
Moreover, the observed backward flow should be 
a net flow of the forward and backward current components and 
the backward components should be inherent among other regions. 
We should emphasize that 
such  backward flow can be obtained, 
only with the channel decomposition method given in the present paper.

\section{Conclusion}\label{conclusion}
An eigen-channel analysis method has been extended to general transport 
quantities based on the nonequilibrium Green's function formalism and 
we studied the transport properties of rodlike nanometer-scale 
metal wires from the viewpoint of ``ill-contact'' in d-orbital channels. 
The ill-contact generates two different characteristics in transport 
phenomena of d-orbital channels; 
the evanescent wave modes and nonlinear conductance at a finite bias voltage. 
Although the present transport calculation has been carried out 
only in cases with model structures, 
the channel decomposition method is general and 
applicable to realistic materials whose atomic structure 
is determined by an energy functional with an electronic structure. 
Such combined simulations of structural and transport properties 
during the formation of nanoscale structures 
should be a foundation of nanoelectronics, 
together with combined experiments of structural and transport observations.

\section*{Acknowledgments} 
Computation was partially carried out using 
the facilities of the Supercomputer Center, Institute for Solid State Physics, 
The University of Tokyo.  
This work was partially supported by a Grant-in-Aid for Scientific Research in Priority
Areas ``Development of New Quantum Simulators and Quantum Design'' (No. 170640004) 
from the Ministry of Education, Culture, Sports, Science, and Technology, Japan. 
H.S. thanks JSPS for the financial support.


\end{document}